\documentstyle[epsf,a4]{article}  
\begin{document}                
\newcommand{\manual}{rm}        
\newcommand\bs{\char '134 }     
\newcommand{\Het}{$^3{\mathrm{He}}$}
\newcommand{\Hef}{$^4{\mathrm{He}}$}
\newcommand{\A}{{\mathrm{A}}}
\newcommand{\D}{{\mathrm{D}}}
\newcommand{\simlt}{\stackrel{<}{{}_\sim}}
\newcommand{\simgt}{\stackrel{>}{{}_\sim}}
\newcommand{\MeV}{\;\mathrm{MeV}}
\newcommand{\TeV}{\;\mathrm{TeV}}
\newcommand{\GeV}{\;\mathrm{GeV}}
\newcommand{\eV}{\;\mathrm{eV}}
\newcommand{\cm}{\;\mathrm{cm}}
\newcommand{\s}{\;\mathrm{s}}
\newcommand{\sr}{\;\mathrm{sr}}
\newcommand{\lab}{\mathrm{lab}}
\newcommand{\ts}{\textstyle}
\newcommand{\ol}{\overline}
\newcommand{\be}{\begin{equation}}
\newcommand{\ee}{\end{equation}}
\newcommand{\ba}{\begin{eqnarray}}
\newcommand{\ea}{\end{eqnarray}}
\newcommand{\rau}{\rho_{\mathrm Au}}
\newcommand{\nn}{\nonumber}
\newcommand{\pp}{$\overline{p}(p)-p\;$}
\renewcommand{\floatpagefraction}{1.}
\renewcommand{\topfraction}{1.}
\renewcommand{\bottomfraction}{1.}
\renewcommand{\textfraction}{0.}               
\renewcommand{\thefootnote}{F\arabic{footnote}}
\title{Related Power-law Growth of Particle
Multiplicities near Midrapidity in Central Au$+$Au
Collisions and in $\ol{p}(p)-p$ Collisions}
\author{Saul Barshay and Georg Kreyerhoff\\
III. Physikalisches Institut\\
RWTH Aachen\\D-52056 Aachen\\Germany}
\maketitle
\begin{center}
To be published in Nucl.~Phys.~A
\end{center}
\begin{abstract}                
A simple power-law growth of charged-particle multiplicities
near midrap\-idity in central Au$+$Au collisions at $\sqrt{s_{NN}}=56$
and 130 GeV, recently measured at RHIC, is derived. We give
predictions for the central particle densities up to $\sqrt{s_{NN}}=1800$
GeV. A strong growth of the Au$+$Au densities above those for 
$\ol{p}(p)-p$ collisions is predicted.
\end{abstract}
Charged-particle multiplicity densities near mid\-rapidity in
Au$+$Au collisions at $\sqrt{s_{NN}}=56$ and 130 GeV are among
the first results\cite{ref1} from the Relativistic Heavy-Ion Collider
(RIHC) at Brookhaven National Laboratory. The most striking feature
of the data is the strong rise of the density (normalized to
the measured number of participating pairs of nucleons) in going
from 56 to 130 GeV, an increase of about 31\%. As seen in Fig.~1,
the rising curve through the data for Au$+$Au collisions is much
steeper than the curve through the corresponding densities measured for
$\ol{p}(p)-p$ collisions over the range of high c.~m.~collision energies,
22 to 1800 GeV\cite{ref2,ref3,ref4,ref5}. The latter curve is steeper
than growth as $\ln s$. In this paper, we show that a simple power-law
growth represents the $\ol{p}(p)-p$ central-density data accur\-ately. The power
is slowly increasing; the values are the same as the powers which have been 
theoretically calculated on the basis of dynamical arguments in
an early study \cite{ref6} of the average charged-particle
multiplicities as a function of $\sqrt{s}$. This is the curve
through the $\ol{p}-p$ data points shown in Fig.~1. Using this power-law
growth, we show that a power-law growth of the density 
holds for central Au$+$Au collisions, and we give
explicit values for the slowly-increasing powers in terms of those
calculated for \pp collisions\cite{ref6}. The power law growth fits
the new data\cite{ref1} for central Au$+$Au collisions, and allows defin\-ite
predictions for higher $\sqrt{s_{NN}}$. These are given for $\sqrt{s_{NN}}$
in the energy domain 200 to 1800 GeV, as shown by the curve through
the Au$+$Au data in Fig.~1. We note that Pb$+$Pb data\cite{ref7} at
$\sqrt{s_{NN}}=17.8$ GeV is also represented by the curve. It is worth 
recalling that general theoretical arguments exist for simple
power-law growth of multiplicities, with a limiting power approached 
asymptotically, in particular the arguments of Polyakov\cite{ref8}
from the early days of approximate scale-invariance and self-similar
processes. Also, power-law (inverse) behaviour has been derived\cite{ref9}
for the asymptotic S-matrix amplitude whose integration over impact parameter
determines the growth of the total \pp cross section at very high 
energies\cite{ref10}.\par
The growth of the normalized particle density near central pseudorapidity,
for \pp collisions, is given by the form,
\be
\frac{dn}{d\eta} =\rho(\sqrt{s}) = {\cal{A}} (\sqrt{s})^{2p(s)}
\ee
where $\cal{A}$ is a constant of order of unity\footnote{The physical meaning of this simple normalization
is about one particle per unit of rapidity at $\sqrt{s}\cong 5$ GeV.
\cite{ref6}}, and
$p(s)$ is a small power which grows slowly with increasing $s$, 
approaching a value of the order of $0.1$ at very high $\sqrt{s}$.
This follows from an accurate representation of the average charged-particle
multiplicities given by the form\cite{ref6},
\be
\langle n(s)\rangle = A f(s) (\ln s) (\sqrt{s})^{2p(s)}
\ee
Here, $A$ is a normalization constant of order unity\cite{ref6}$^{F1}$,
$f(s)$ is a function which approaches approximately unity at large $s$, and
the $(\ln s)$ factor represents the increasing extent of the rapidity ``plateau''.
Thus the essential dependence is as $(\ln s)(\sqrt{s})^{2p(s)}$.
The slowly increasing $p(s)$ have been theoretically calculated at
12 values of $\sqrt{s}$ from 14 to 40,000 GeV (as given in Table I
of \cite{ref6}). The calculation involves a single phenomenological
parameter which controls the (slowing) growth of $p(s)$ toward
a limiting value of order of $0.1$ (Eq.~(5) in \cite{ref6}). The
calculation is based upon the physical idea that collision energy
which is progressively removed from the extreme fragmentation region
sustains a power-law growth of $\langle n(s)\rangle$ with a 
slowly increasing power, i.~e.~ an increasingly high multiplicity of
particles over the rapidity plateau. This is consistent with the entire
central \pp collision system becoming ``blacker'' over the 
impact-parameter plane.\cite{ref9}\footnote{It is noteworthy that analyses of certain characteristics of 
cosmic-ray air showers above $10^{18}$ eV suggest a more strongly growing hadronic
cross section and/or ``inelasticity''.} We assume that the
previously calculated powers, $p(s)$ used in Eq.~(2),
are approximately the powers to be used in Eq.~(1). Using the
calculated $p(s)$ from Table I in \cite{ref6}, we obtain from Eq.~(1)
the following $\rho(\sqrt{s})$, for ${\cal{A}}=1.2$ (with $\sqrt{s}$
in GeV),
\ba 
pp & \sqrt{s}=22 & p(s)\cong(0.043-0.055) \rightarrow \rho(22) \cong (1.6-1.7)\nn\\
\ol{p}p & \sqrt{s}=53, & p(s)\cong 0.055 \rightarrow \rho(53) \cong 1.9\nn\\
& \sqrt{s}=200, & p(s)\cong 0.069 \rightarrow \rho(200) \cong 2.5\nn\\
& \sqrt{s}=546, & p(s)\cong 0.075 \rightarrow \rho(546) \cong 3.1\\
& \sqrt{s}=630, & p(s)\cong 0.075 \rightarrow \rho(630) \cong 3.2\nn\\
& \sqrt{s}=900, & p(s)\cong 0.078 \rightarrow \rho(900) \cong 3.45\nn\\
& \sqrt{s}=1800, & p(s)\cong 0.081 \rightarrow \rho(1800)\cong 4\nn
\ea
The resulting curve shown in Fig.~1 is an accurate representation of
the data. Eq.~(1) gives agreement with 7 explicit data points with a
single parameter\cite{ref6} controlling the growth of the
power $p(s)$, and the normalization $\cal{A}$. \par
Given the basic power-law behavior, then for central Au$+$Au collisions,
the growth of the normalized particle density near midrapidity
can also be given approximately by a simple
power-law form, with \underline{known powers} and normalization,
\ba
\frac{1}{0.5\langle N_{\mathrm{part}}\rangle}\frac{dN}{d\eta}&=&
\rho_{\mathrm Au} (\sqrt{s_{NN}})\nn\\ 
&\cong& \left\{
{\cal{A}}(\sqrt{s_{NN}})^{2p(s_{NN})}\right\}\times\left\{
{\cal{A}}\left(\frac{\sqrt{s_{NN}}}{\langle n(s_{NN})\rangle}\right)^{2\tilde{p}}
\right\}
\ea
The first quantity in brackets represents the multiplicity from each 
of $0.5\langle N_{\mathrm part}\rangle$ collisions\cite{ref1} where
$\langle N_{\mathrm part}\rangle$ is the experimentally estimated\cite{ref1}
average number of participants at $\sqrt{s_{NN}}$. The second quantity
in brackets gives rise to an expanded multiplicity, in the approximation
in which each particle from the initial multiplicity undergoes an
additional collision\footnote{We approximate the multiplicity growth in the same way
for hadron-hadron collisions i.~e.~pion-pion, pion-nucleon,
nucleon-nucleon. At high enough energies, one might make explicit
reference to partonic collisions. The notion of partonic "saturation"
at very high $\sqrt{s}$ can be incorporated in the limiting value
of the power $p(s)$.\cite{ref9} The last factor in Eq.~(4) can be generalized
to include neutrals, and a lower, effective $\sqrt{s}$ for centrally-produced particles.
As an example, a resulting approximate numerical factor of $\sim 1.5\left(\frac{1}{1.5}
\times \frac{1}{20}\right)^{0.1}$ is close to unity. Eq.~(4) can be generalized to
$A(\sqrt{s})^{2p(s)}\left\{(1-x)+xA(\sqrt{s}/\langle n(s)\rangle)^{2\tilde{p}}\right\}$,
where the function $x\cong 0$ for \pp collisions goes continuously to $x\cong 1$ for
central collisions of two heavy ions, at present $\sqrt{s}$.} with a typical collision energy
approximated roughly as $\sqrt{\tilde{s}_{NN}}=(\sqrt{s_{NN}}/
\langle n(s_{NN})\rangle)$. This gives rise to multiplication of the
initial multiplicity by the bracketed number which is determined by the
power $\tilde{p}$, the value of $p(s)$ appropriate to $\sqrt{\tilde{s}_{NN}}$.
Additional collisions can occur, but in the energy range $\sqrt{s_{NN}}$
from 56 to 1800 GeV discussed below, further reduction to a typical
subsequent collision energy results in low energy. In calculating 
$\rau(\sqrt{s_{NN}})$ from Eq.~(4), we use the values of the average multiplicity
$\langle n(\sqrt{s_{NN}})\rangle$ as these are tabulated in Table I 
of \cite{ref6}. We give the effective power \footnote{
In our calculations with all relevant $\sqrt{\tilde{s}_{NN}}
<50$ GeV, we use an approximate $\tilde{p}(\sqrt{\tilde{s}_{NN}})
= \tilde{p}(53)=0.055$.\cite{ref6} In more detail, use of smaller powers
at low energies\cite{ref6} would tend to be compensated
by use of higher values of effective energy for some collisions 
i.~e.~nucleon-nucleon, valence quarks.$^{F3}$} of $s_{NN}$,
$p'=(p(s_{NN})+\tilde{p}(\tilde{s}_{NN}))$ and the results for 
$\rau(\sqrt{s_{NN}})$,
\ba
& \sqrt{s_{NN}}=56,& p'\cong 0.11 \rightarrow \rau(56)\cong 2.6\nn\\
& \sqrt{s_{NN}}=130,& p'\cong 0.12 \rightarrow \rau(130)\cong 3.4\nn\\
& \sqrt{s_{NN}}=200,& p'\cong 0.125 \rightarrow \rau(200)\cong 4\\
& \sqrt{s_{NN}}=1800,& p'\cong 0.135 \rightarrow \rau(1800)\cong 7.1\nn
\ea
As a check, for Pb$+$Pb collisions at a relatively low \footnote{
Then $(\sqrt{s_{NN}}/\langle n(s_{NN})\rangle)$ is $\simlt 3$
GeV.} 
$\sqrt{s_{NN}}=17.8$, $p'\cong 0.11 \rightarrow \rho_{\mathrm Pb}(17.8)
\cong 2$.\cite{ref7} The resulting curve in Fig.~1 represents the present
the data and predicts the densities at higher $\sqrt{s_{NN}}$,
from 200 to 1800 GeV. With the measured\cite{ref1} $\langle N_{\mathrm part}
\rangle = 330$ at 56 GeV and 343 at 130 GeV, our calculated values of 
$(dN/d\eta)$ are 430 and 580, respectively. The experimental 
numbers for $(dN/d\eta)_{|\eta<1|}$ \cite{ref1} are respectively, 
$408 \pm 12(\mathrm{stat}) \pm 30 (\mathrm{syst})$, and
$555 \pm 12(\mathrm{stat}) \pm 35 (\mathrm{syst})$.
The predicted $(dN/d\eta)$ are 680 at 200 GeV (for $\langle N_{\mathrm part}
\rangle = 343$), and 1220 (1350) at 1800 GeV (for  $\langle N_{\mathrm part}
\rangle = 343 (380) $). In Eq.~(4), the quantity $(\sqrt{s_{NN}}/
\langle  n(s_{NN})\rangle)^{2\tilde{p}}$ gives an increase in the 
multiplicity by about 20\% at 56 GeV. This increases to $\sim$30\% at
130 GeV, and to $\sim$50\% at 1800 GeV. It is useful to compare these
multiplicities with another recent calculation \cite{ref16}, which 
gives a low value of $\sim 945$ at 1800 GeV (i.~e.~$\rau(1800)\cong 5.5$).
This calculation involves applying a large ``correction'', motivated
by speculative dynamics, to reduce excessive multipicities from a dual
model (already necessary at 56 GeV, i.~e.~note Fig.~1 in \cite{ref16}).
This results in a very strong suppression of multiplicities in the 
TeV range\cite{ref17}. \par
In conclusion, although scale invariance \cite{ref8} is not exact, 
a global \footnote{
The densities as such, make no direct reference to
particle momenta. }
 collision property like particle densities near
midrapidity for Au$+$Au collisions and \pp collisions
can be directly related and quantitatively represented by a 
physically motivated, simple
power-law growth with energy.\footnote{
Note that $\ol{p}-p$ data is parameterized in 
\cite{ref5} and \cite{ref16} by very different sets of the 3
parameters in the form $A+B(\ln s) + C(\ln s)^2$, without 
physical motivation.}
 This is relevant to simply estimating
the maximum energy density\cite{ref1}, which is reached only for
a very brief collision time.
\section*{Added note}
Since completion of this paper, new results have appeared from RHIC
in this rapidly developing field. First, data on the ``centrality''
dependence for Au$+$Au collisions at $\sqrt{s_{NN}}=130\GeV$. i.~e.~$\rho_{\mathrm{Au}}$
versus $\langle N_{\mathrm{part}}\rangle$ appears in Fig.~4 of \cite{ref13}.
This can be approximately, but simply described by the generalized formula
in footnote F3. At $\sqrt{s_{NN}} = 130\GeV$, write $\rho\cong
(2.25)\left\{ (1-x)+x(1.2)(1.3)\right\}$, with the parameterization
$x=a\ln(N_P/b) \cong \ln(N_P/10)$, where $N_P=\langle N_{\mathrm{part}}\rangle$.
Then,
\ba
\rho_{\mathrm{Au}}\cong 2.25 &{\mathrm{at}} & N_P \cong 10\;\;\;(x\to \sim 0)\nn\\
\rho_{\mathrm{Au}}\cong 2.75 &{\mathrm{at}} & N_P \cong 40\nn\\
\rho_{\mathrm{Au}}\cong 3.05 &{\mathrm{at}} & N_P \cong 100\nn\\
\rho_{\mathrm{Au}}\cong 3.3 &{\mathrm{at}} & N_P \cong 200\nn\\
\rho_{\mathrm{Au}}\cong 3.45 &{\mathrm{at}} & N_P \cong 300\nn\\
\rho_{\mathrm{Au}}\cong 3.5 &{\mathrm{at}} & N_P \cong 350\;\;\;(x\to \sim 1)\nn
\ea
We assume that $x$ remains $\sim 1$ for $N_P\ge 350$ at $\sqrt{s_{NN}}>130\GeV$.
It is noteworthy that $d\rho_{\mathrm{Au}}/dN_P \propto 1/N_P$, is similar
to the $d\rho_{\mathrm{Au}}/dN_P\propto 1/N_P^{0.84}$ obtained from the
phenomenological fit $\rho_{\mathrm{Au}}N_P\propto N_P^{(1.16\pm 0.04)}$
quoted in \cite{ref13}. The above representation of $\rho_{\mathrm{Au}}(N_P)$
does not necessarily require a markedly increased fraction of ``hard''
processes at $\sqrt{s_{NN}}=130\GeV$. In this connection, there is new
data on the transverse energy density $dE_T/d\eta$, which indicates
\cite{ref14} that $(dE_T/d\eta)(dN_{ch}/d\eta)^{-1}$ does not increase
markedly in going down from $\sqrt{s_{NN}}\cong 17\GeV$ to 130 GeV.
The present ideas are consistent with this behavior, because the energy
is going mainly into a marked increase in particle production
(relative to $\ol{p}(p)-p$), not into a substantially larger fraction of ``hard''
processes.\par
Finally, with respect to Fig.~1, we note that a recent measurement\cite{ref15}
for Pb$+$Pb at $\sqrt{s_{NN}}\cong 17\GeV$, when scaled to Au$+$Au, gives a higher
value $\rho_{\mathrm{Au}}\sim 2.46$. Also, we note that our curve for $p\ol{p}$
reaches the value of $\rho\sim 1.6$ at $\sqrt{s}=22 GeV$, near to the value
measured for $pp$-collisions\cite{ref2}.\par
We thank Dr.~Mark Tannenbaum of the PHENIX collaboration at RHIC for
stimulating correspondence.

\newpage
\section*{Figures}
\begin{figure}[b]
\begin{center}
\mbox{\epsfysize 12cm\epsffile{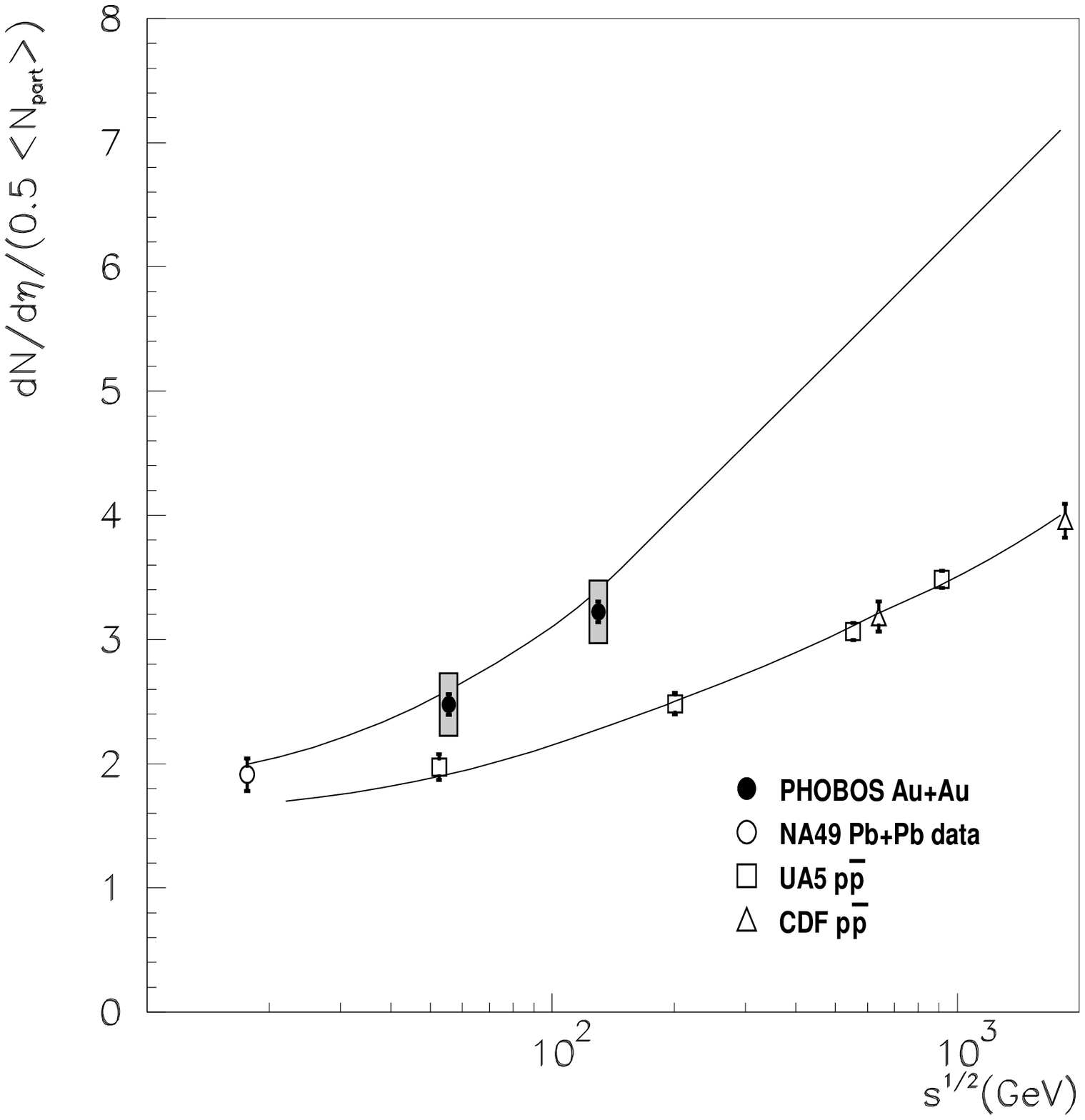}}
\caption{Measured pseudorapidity density near midrapidity normalized per measured 
participant pair, for central Au$+$Au collisions at $\sqrt{s_{NN}}=56$ and
130 GeV.\cite{ref1} The point at 17.8 GeV is for Pb-Pb collisions.\cite{ref7}
The predicted curve is calculated from Eq.~(4). For comparision, 
$\ol{p}-p$ data points from \cite{ref3,ref4,ref5} are shown. This curve is
from Eq.~(1). Both curves are simple power laws with related, 
theoretically-calculated powers\cite{ref6}, and related
normalization.}
\end{center}
\end{figure}

\begin{thebibliography}{99}
\bibitem{ref1}PHOBOS Collab., B.~B.~Back et. al., Phys.~Rev.~Lett.
{\bf 85}, (2000) 3100.
\bibitem{ref2}NA22 Collab., M.~Adamus et al., Z.~Phys.~{\bf C37}, (1988) 215.
\bibitem{ref3}UA5 Collab., G.~J.~Alner et. al., Phys.~Rep.~{\bf 154}, (1987) 247.
\bibitem{ref4}UA5 Collab., R.~E.~Ansorge et al., Z.~Phys.~{\bf C43}, (1989) 357.
\bibitem{ref5}CDF Collab., F.~Abe et al., Phys.~Rev.~{\bf D41}, (1990) 2330.
\bibitem{ref6}S.~Barshay and P.~Heiliger, Phys.~Rev.~{\bf D47}, (1993) 4150.
\bibitem{ref7}J.~B\"achler et al., Nucl.~Phys.~{\bf A661}, (1999) 45.
\bibitem{ref8}A.~M.~Polyakov, Zh.~Eksp.~Teor.~Fiz. {\bf 59}, (1970) 542
[ Sov.~Phys.~JETP {\bf 32}, (1971) 296].
\bibitem{ref9}S.~Barshay, P.~Heiliger, and D.~Rein, 
Mod.~Phys.~Lett.~{\bf A7}, (1992) 2259. This paper also contains a
previously unpublished derivation due to R.~P.~Feynman.
\bibitem{ref10} S.~Barshay, P.~Heiliger, and D.~Rein, Z.~Phys.~{\bf C52}, (1991) 415; 
Z.~Phys.~{\bf C56}, (1992) 77.
\bibitem{ref16} J.~Dias de Deus and R.~Ugoccioni, Phys.~Lett.~{\bf B491}, (2000) 
253.
\bibitem{ref17} C.~Pajares, D.~Sousa, and R.~A.~V\'azquez, 
Phys.~Rev.~Lett.~{\bf 86}, (2001) 1674.
\bibitem{ref13} PHENIX Collab., K.~Adcox et al., Phys.~Rev.~Lett.~{\bf 86} (2001) 3500.
\bibitem{ref14} PHENIX Collab., K.~Adcox et al., nucl-ex/0104015.
\bibitem{ref15} WA98  Collab., M.~M.~Aggarwal et al., Eur.~Phys.~J.~{\bf C18} (2001) 651.
\end{thebibliography}
\end{document}